\documentclass{article}

\usepackage{arxiv}

\usepackage[utf8]{inputenc} 
\usepackage[T1]{fontenc}    
\usepackage{hyperref}       
\usepackage{url}            
\usepackage{booktabs}       
\usepackage{amsfonts}       
\usepackage{nicefrac}       
\usepackage{microtype}      
\usepackage{lipsum}		
\usepackage{graphicx}
\usepackage{natbib}
\usepackage{doi}
\usepackage{stfloats}
\usepackage{multirow}
\usepackage{makecell} 
\usepackage{xcolor}

\title{Leveraging large language models for SQL behavior-based database intrusion detection}

\author{ 
  \href{https://orcid.org/0009-0000-7555-7506}{\includegraphics[scale=0.06]{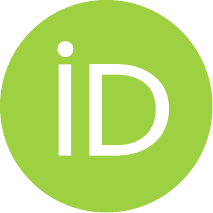}\hspace{1mm}Meital Shlezinger}\thanks{Corresponding author.} \\
  Huawei Tel Aviv Research Center \\
  Tel Aviv, Israel \\
  \texttt{mshlezinger@gmail.com} \\
  \And
  Shay Akirav \\
  Huawei Tel Aviv Research Center \\
  Tel Aviv, Israel \\
  \And
  Lei Zhou \\
  Huawei Tel Aviv Research Center \\
  Tel Aviv, Israel \\
  \And
  Liang Guo \\
  Huawei Technologies Co. LTD Shenzhen \\
  Guangdong, China \\
  \And
  Avi Kessel \\
  Huawei Tel Aviv Research Center \\
  Tel Aviv, Israel \\
  \And
  Guoliang Li \\
  Tsinghua University \\
  Beijing, China \\
}

\renewcommand{\shorttitle}{\textit{arXiv} Template}

\hypersetup{
pdftitle={A template for the arxiv style},
pdfsubject={q-bio.NC, q-bio.QM},
pdfauthor={David S.~Hippocampus, Elias D.~Striatum},
pdfkeywords={First keyword, Second keyword, More},
}

\begin{document}
\maketitle

\begin{abstract}
Database systems are extensively used to store critical data across various domains. However, the frequency of abnormal database access behaviors, such as database intrusion by internal and external attacks, continues to rise. Internal masqueraders often have greater organizational knowledge, making it easier to mimic employee behavior effectively. In contrast, external masqueraders may behave differently due to their lack of familiarity with the organization. Current approaches lack the granularity needed to detect anomalies at the operational level, frequently misclassifying entire sequences of operations as anomalies, even though most operations are likely to represent normal behavior. On the other hand, some anomalous behaviors often resemble normal activities, making them difficult for existing detection methods to identify. This paper introduces a two-tiered anomaly detection approach for Structured Query Language (SQL) using the Bidirectional Encoder Representations from Transformers (BERT) model, specifically DistilBERT, a more efficient, pre-trained version. Our method combines both unsupervised and supervised machine learning techniques to accurately identify anomalous activities while minimizing the need for data labeling. First, the unsupervised method uses ensemble anomaly detectors that flag embedding vectors distant from learned normal patterns of typical user behavior across the database (out-of-scope queries). Second, the supervised method uses fine-tuned transformer-based models to detect internal attacks with high precision (in-scope queries), using role-labeled classification, even on limited labeled SQL data. Our findings make a significant contribution by providing an effective solution for safeguarding critical database systems from sophisticated threats.
\end{abstract}


\section{Introduction}
In recent years, database security has made significant advancements, driven by the increasing reliance on databases and the threat of targeted attacks. Data breaches and malicious activities have raised significant security and privacy concerns for organizations, emphasizing the critical need for robust database security. The 2023 Tesla data leak, which compromised the personal data of 75,000 of the company's employees, is a stark reminder of these vulnerabilities~\cite{cyfor2023}. Tesla attributed the breach to 2 former employees, highlighting the importance of implementing more effective security measures. Protecting databases from leakage of sensitive information requires addressing internal and external threats. Insider threats occur when legitimate system users abuse their access privileges to steal or leak sensitive data. In contrast, external threats originate from attackers outside the organization who exploit network or system vulnerabilities to gain access to the organization. Insider threats are particularly concerning, as they involve trusted individuals who have permission to access various data and services. According to the Ponemon Institute’s 2023 report ``The Cost of Insider Risk,'' 71\% of the companies surveyed reported experiencing between 21 and over 40 insider incidents annually, a 4\% increase from 2022~\cite{ponemon2023}.

\begin{figure*}[t]
  \centering
  \includegraphics[width=\linewidth]{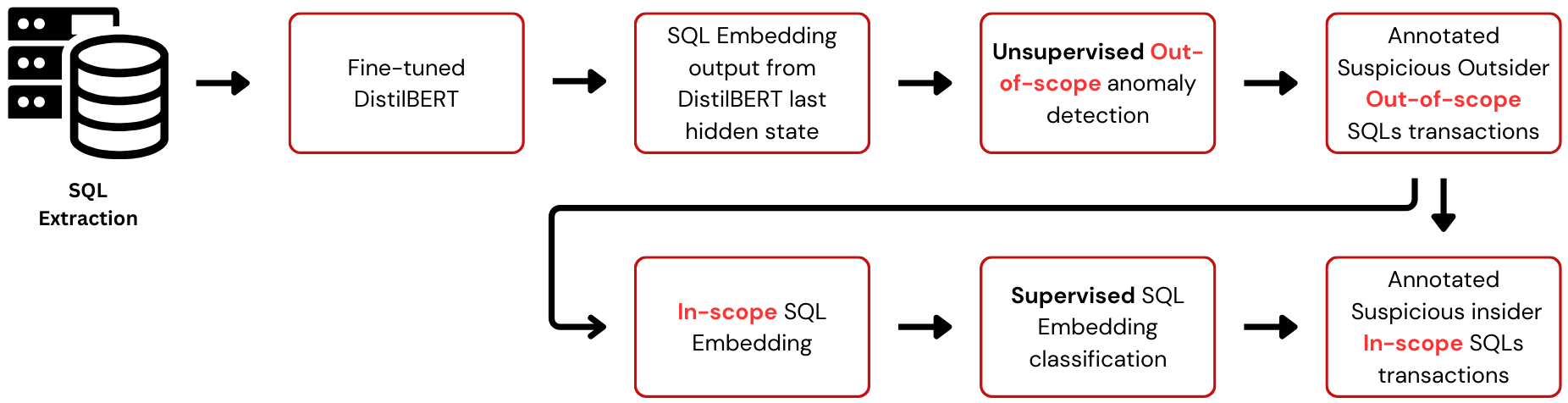}
  \caption{An overview of a two-tier anomaly detection framework composed of unsupervised and supervised approaches for external (out-of-scope queries) and internal (in-scope queries) database attacks.}
  \label{Fig1}
\end{figure*}

Traditional security measures, such as authentication, role-based access control, data encryption, and physical security, provide a foundational level of protection. However, protecting databases from legitimate system users abusing their access privileges poses a continuous challenge, emphasizing the need for effective security controls to mitigate insider threats. One particularly concerning type of attack is a masquerade attack, where an attacker uses stolen credentials to impersonate a legitimate employee and gain unauthorized access to resources, including databases. Masquerade attacks can occur in 2 ways: (i) an insider gains control of another employee's credentials with a different privilege level, or (ii) an outsider obtains a legitimate employee's credentials. Detecting such attacks requires specialized techniques like masquerade anomaly detection~\cite{bertacchini2008survey}. This approach aims to identify unauthorized users by analyzing deviations in user behavior between transactions, which may indicate an attacker's presence.

\begin{figure*}[b]
  \centering
  \includegraphics[width=\linewidth]{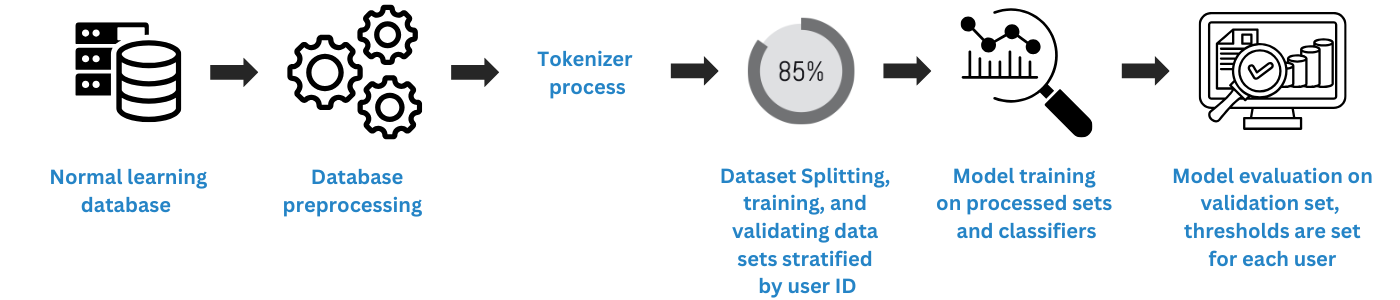}
  \caption{Supervised method – Learning period flow chart.}
  \label{Fig2}
\end{figure*}

A significant body of literature has been dedicated to anomaly detection in database systems, particularly within Structured Query Language (SQL). These detection methods can be divided into 3 primary categories: syntax-based techniques~\cite{hussain2015detanom, sallam2017dbsafe}, context-based methods~\cite{alizadeh2018behavior, gafny2011poster, khan2017detecting, wu2009database}, and data statistics-based approaches~\cite{khan2018semantic,mathew2010data}. Despite their contributions, these methods have notable limitations. One major drawback is their inability to fully capture SQL's underlying structure and syntax, often resulting in a relatively high rate of false positives.
Natural Language Processing (NLP) has introduced new possibilities for addressing complex security challenges. In particular, transformer-based models, such as Bidirectional Encoder Representations from Transformers (BERT)~\cite{devlin2019bert}, Large Language Model Meta AI (LLaMA), and LLaMA 2~\cite{touvron2023llama(a), touvron2023llama(b)}, have achieved state-of-the-art (SOTA) performance in various NLP tasks. These models can enhance computer security by enabling the development of more effective and adaptable anomaly detection systems capable of learning from large-scale and diverse data sources.

Detecting insider threats presents considerable challenges, particularly due to limitations in threat data availability and quality. Recent surveys indicate that advanced deep learning models, such as Long Short-Term Memory (LSTM) networks and transformer-based models, can address some of these data issues using techniques like anomaly detection~\cite{guo2021logbert}. Transformer-based models, known for their capability to handle large datasets and manage long-range dependencies through multi-head attention, provide a promising alternative to traditional machine learning models.
DistilBERT~\cite{sanh2019distilbert}, a lighter and more efficient version of BERT~\cite{devlin2019bert}, has shown exceptional performance across various NLP tasks. By fine-tuning a DistilBERT model on a SQL dataset, we can capture the underlying structure and syntax of SQL. This enables the model to detect deviations in users' SQL commands from normal activity with typical patterns. Consequently, it is crucial to assess whether these methods can accurately identify specific anomalous operations and report them to system operators for further action.

\begin{figure*}[t]
  \centering
  \includegraphics[width=\linewidth]{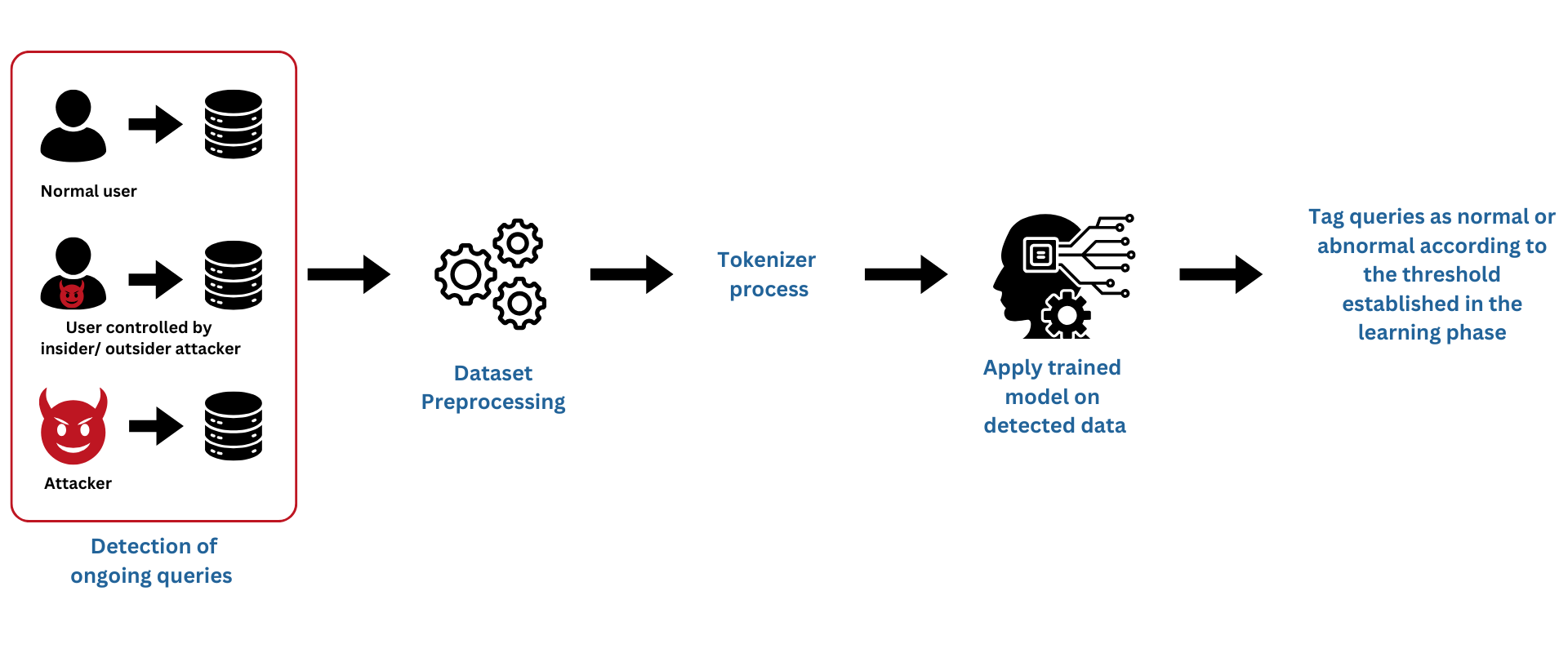}
  \caption{Supervised method - Detection period flow chart.}

  \label{Fig3}
\end{figure*}

Our proposed system transcends the limitations of conventional rule-based approaches by harnessing the adaptive capabilities of machine learning to address evolving threats. This is achieved by applying a Large Language Model (LLM), a deep learning algorithm capable of performing various NLP tasks. LLMs, which use transformer models trained on large datasets, can recognize, translate, predict, or generate text or other content based on the attention and context of words. The proposed system aims to analyze query patterns and user behaviors in real time, enabling the early detection of anomalous activities indicative of malicious intent.
Our current research primarily focuses on employing machine learning techniques to understand the semantics and context of SQL transactions. We aim to accurately detect anomalies and enhance overall system security using LLMs.

The main contributions of this paper are as follows:
1. We implement a two-tier anomaly detection framework for SQL that (i) utilizes the pre-trained DistilBERT model and ensemble anomaly detectors to address a significant issue in database security, including user behavior for external attacks (out-of-scope queries), and (ii) uses a role-labeled classification method to detect internal attacks by transformer-based models (see Fig. 1).
2. A role (user)-labeled classification system leveraging detailed behavioral profiles across several distinct roles (see Figs. 2-3), assuming all queries for all roles are normal in the learning period, while in the detection phase, the system identifies abnormal behavior using probabilistic embedding thresholds. This approach enables fine-grained detection of internal masqueraders, going beyond the binary 'Normal/Abnormal' labeling used in prior work. 
3. In the detection phase, for internal threats, our supervised model identifies anomalies in two ways: (i) when a query is most likely associated with a different user than the user it was labeled with, and (ii) when a query matches the correct user but its probability score falls below the user’s learned threshold (see Section 5.2). This strategy enables the detection of two distinct types of internal masquerade attacks. 
4. We assess the performance of the supervised fine-tuned models on a few-shot set of labeled SQL data, emphasizing the adaptability and accuracy of our method.

Fig.~\ref{Fig1} describes the general pipeline of the two-tier approach to detecting external and internal database attacks by using both unsupervised and supervised approaches. The rest of this paper is structured as follows: Section 2 reviews related work in SQL anomaly detection; Section 3 describes the data, including dataset details and data cleaning methods; Section 4 outlines our methodology, covering both unsupervised and supervised approaches; Section 5 presents the results along with examples of anomaly activities; and Section 6 concludes the paper.

\section{Related Work}
\subsection{Database Anomaly Detection}
Various methods have been suggested for identifying anomalies in databases, which can be categorized into 3 primary approaches: (i) syntax-based techniques, which principally use the syntax of SQL statements to pinpoint anomalies~\cite{sallam2017dbsafe, hussain2015detanom}; (ii) context-based methods, which consider contextual features while modeling and learning patterns of normal behavior to detect deviations~\cite{alizadeh2018behavior, gafny2011poster, khan2017detecting, wu2009database}; and (iii) data statistics-based methods which identify anomalies by observing significant data changes caused by operational behavior~\cite{khan2018semantic, mathew2010data}. Additionally, some hybrid approaches combine elements from these categories (syntax-based, context-based, and data statistics-based) to enhance detection accuracy~\cite{wu2009database, gafny2011poster}.
Despite the advancements in anomaly detection techniques, traditional methods still face significant challenges. Specifically, these challenges include (i) an Inability to capture the complete structure and syntax of the SQL; (ii) difficulties in distinguishing between anomalous and normal behavior, especially when queries are similar but not identical, requiring advanced sentence processing to capture contextual information. For example: ``select $\ast$ from employees where depid = ?''  compared to ``select $\ast$ from employees where depid = ? and managerid = ?'', both queries have similar meaning, but they are not identical; and (iii) limitations in root cause analysis as most approaches fail to differentiate between external and internal attack methods. 

\begin{table*}[h] 
  \caption{Supervised method output - Users Probability Matrix, the highest probability for each row (query) marked in bold red, and the significant probabilities marked in non-bold red, attributed to a specific user (labeled from 0 to 10).}
  \label{Table1}
    \resizebox{1\textwidth}{!}{ 
  \begin{tabular}{rlllllllllll} 
    \toprule
    & 0 & 1 & 2 & 3 & 4 & 5 & 6 & 7 & 8 & 9 & 10 \\
    \midrule
    1 & 0.004128 & 0.003537 & \textbf{\textcolor{red}{0.956105}} & 0.005157 & 0.005614 & 0.005391 & 0.003488 & 0.003129 & 0.00824 & 0.003558 & 0.001616 \\
    2 & 0.006137 & 0.006842 & 0.004977 & 0.003585 & 0.003422 & \textbf{\textcolor{red}{0.936215}} & 0.005186 & 0.00384 & 0.009933 & 0.009418 & 0.010445 \\
    3 & 0.007922 & 0.003449 & 0.017589 & 0.004932 & 0.013318 & 0.008284 & \textcolor{red}{0.922157} & 0.007796 & 0.007715 & 0.003756 & 0.003101 \\
    4 & 0.000132 & 7.41E-05 & 5.91E-05 & 5.37E-05 & 0.000115 & 0.000122 & 0.000145 & 0.000994 & 0.000153 & 0.000468 & \textbf{\textcolor{red}{0.997685}} \\
    5 & 0.000785 & 0.002707 & 0.001228 & \textbf{\textcolor{red}{0.988784}} & 0.001185 & 0.001504 & 0.000663 & 0.000595 & 0.001566 & 0.000676 & 0.000307 \\
    6 & 0.000651 & 0.000365 & 0.000291 & 0.000265 & 0.000565 & 0.000601 & 0.000715 & 0.004897 & 0.000756 & \textcolor{red}{0.933976} & 0.056918 \\
    7 & 0.001395 & 0.0005 & 0.000629 & 0.000567 & 0.001362 & 0.001459 & 0.001522 & \textcolor{red}{0.981272} & 0.002977 & 0.002176 & 0.006141 \\
    8 & \textcolor{red}{0.966004} & 0.011226 & 0.004079 & 0.006021 & 0.007439 & 0.000944 & 0.001278 & 0.001049 & 0.001038 & 0.000505 & 0.000417 \\
    9 & 0.004952 & 0.003528 & \textcolor{red}{0.942357} & 0.005695 & 0.009944 & 0.009494 & 0.004185 & 0.003753 & 0.009885 & 0.004268 & 0.001939 \\
    10 & 0.008037 & 0.002881 & 0.004841 & 0.0041 & 0.013512 & 0.008404 & \textbf{\textcolor{red}{0.935533}} & 0.007909 & 0.007827 & 0.003811 & 0.003146 \\
    11 & 0.009278 & 0.004265 & 0.008231 & 0.004734 & 0.008598 & 0.009702 & 0.00841 & 0.021829 & \textcolor{red}{0.835873} & 0.08545 & 0.003632 \\
    12 & 0.001284 & 0.00072 & 0.000574 & 0.000521 & 0.001114 & 0.001184 & 0.001409 & \textbf{\textcolor{red}{0.98216}} & 0.00149 & 0.003892 & 0.005652 \\
    13 & 0.001291 & 0.000463 & 0.001145 & 0.000658 & 0.00217 & 0.00135 & 0.001408 & \textcolor{red}{0.980904} & 0.002754 & 0.002013 & 0.005846 \\
    14 & 0.002027 & \textcolor{red}{0.80957} & 0.007335 & 0.164282 & 0.003062 & 0.003887 & 0.001713 & 0.001536 & 0.004047 & 0.001747 & 0.000794 \\
    15 & 0.00041 & 0.000178 & 0.000501 & 0.000203 & 0.0004 & 0.000428 & 0.000447 & 0.003106 & 0.000874 & 0.000692 & \textcolor{red}{0.99276} \\
    16 & 0.00041 & 0.00023 & 0.000183 & 0.000166 & 0.000355 & 0.000378 & 0.00045 & 0.003106 & 0.000475 & 0.001581 & \textcolor{red}{0.992665} \\
    17 & 0.002728 & 0.003041 & 0.001954 & \textcolor{red}{0.778131} & 0.001521 & 0.198728 & 0.002305 & 0.001922 & 0.004415 & 0.004186 & 0.001068 \\
    18 & 0.00041 & 0.000178 & 0.000501 & 0.000203 & 0.0004 & 0.000428 & 0.000447 & 0.003106 & 0.000874 & 0.000692 & \textcolor{red}{0.99276} \\
    19 & 0.000132 & 7.41E-05 & 5.91E-05 & 5.37E-05 & 0.000115 & 0.000122 & 0.000145 & 0.000994 & 0.000153 & 0.000467 & \textbf{\textcolor{red}{0.997685}} \\
    20 & 0.010387 & \textbf{\textcolor{red}{0.913162}} & 0.006296 & 0.008674 & 0.09438 & 0.00811 & 0.008777 & 0.007319 & 0.00783 & 0.01594 & 0.004066 \\
    21 & 0.00041 & 0.00023 & 0.000183 & 0.000166 & 0.000355 & 0.000378 & 0.00045 & 0.003106 & 0.000475 & 0.001448 & \textcolor{red}{0.992798} \\
    22 & 0.006917 & 0.00388 & 0.003094 & 0.002809 & 0.006 & 0.006378 & 0.00759 & \textcolor{red}{0.943863} & 0.003674 & 0.013089 & 0.002708 \\
    23 & 0.11758 & 0.13109 & \textcolor{red}{0.867791} & 0.012224 & 0.014204 & 0.021015 & 0.009935 & 0.008286 & 0.019032 & 0.018044 & 0.004603 \\
    24 & 0.000891 & 0.000319 & 0.000402 & 0.000362 & 0.00087 & 0.000931 & 0.000971 & 0.006752 & 0.0019 & \textcolor{red}{0.963781} & 0.022821 \\
    25 & 0.004518 & 0.001619 & 0.002038 & 0.002241 & \textbf{\textcolor{red}{0.967406}} & 0.004001 & 0.00542 & 0.004446 & 0.0044 & 0.002142 & 0.001769 \\
    26 & 0.000643 & 0.002218 & 0.003479 & \textcolor{red}{0.988729} & 0.000971 & 0.00084 & 0.000543 & 0.000487 & 0.001284 & 0.000554 & 0.000252 \\
    27 & \textcolor{red}{0.967515} & 0.011243 & 0.002522 & 0.006031 & 0.007451 & 0.000945 & 0.00128 & 0.00105 & 0.001039 & 0.000506 & 0.000418 \\
    28 & 0.00041 & 0.0003 & 0.000496 & 0.000203 & 0.000355 & 0.000378 & 0.000449 & 0.003105 & 0.000475 & 0.001447 & \textcolor{red}{0.992381} \\
    29 & 0.000132 & 7.41E-05 & 5.91E-05 & 5.37E-05 & 0.000115 & 0.000122 & 0.000145 & 0.000994 & 0.000153 & 0.000467 & \textbf{\textcolor{red}{0.997685}} \\
    30 & 0.004498 & 0.001959 & 0.005498 & 0.002723 & \textcolor{red}{0.96324} & 0.03983 & 0.005396 & 0.004427 & 0.004381 & 0.002133 & 0.001761 \\
    31 & 0.004637 & 0.001662 & 0.002793 & 0.00289 & \textcolor{red}{0.966278} & 0.004106 & 0.003918 & 0.003514 & 0.004516 & 0.003871 & 0.001815 \\
    32 & 0.00041 & 0.0003 & 0.000496 & 0.000203 & 0.000355 & 0.000378 & 0.000449 & 0.003105 & 0.000475 & 0.001477 & \textcolor{red}{0.992381} \\
    33 & 0.00017 & 0.00019 & 0.000122 & 9.94E-05 & 9.49E-05 & 0.000779 & 0.000154 & 0.001029 & 0.005801 & \textbf{\textcolor{red}{0.987493}} & 0.004066 \\
    34 & 0.000149 & 8.35E-05 & 6.66E-05 & 6.05E-05 & 0.000107 & 0.000137 & 0.000135 & 0.0009 & 0.000173 & \textcolor{red}{0.985179} & 0.01301 \\
    35 & 0.001571 & 0.001752 & 0.001125 & 0.000918 & 0.000876 & 0.007192 & 0.001424 & 0.002848 & \textcolor{red}{0.950879} & 0.030799 & 0.000615 \\
    36 & 0.000528 & 0.000189 & 0.000238 & 0.000215 & 0.000516 & 0.000552 & 0.000576 & 0.004005 & 0.001127 & 0.000892 & \textcolor{red}{0.991161} \\
    37 & \textbf{\textcolor{red}{0.974423}} & 0.011324 & 0.000657 & 0.003961 & 0.004358 & 0.000952 & 0.001289 & 0.001058 & 0.001047 & 0.00051 & 0.000421 \\
    38 & 0.000221 & 0.000157 & 0.000159 & 0.013857 & 0.000139 & 0.001146 & 0.0002 & 0.001334 & 0.010568 & \textcolor{red}{0.966562} & 0.005657 \\
    39 & 0.001393 & 0.000607 & 0.001703 & 0.00069 & 0.001361 & 0.001457 & 0.00152 & \textcolor{red}{0.97999} & 0.002973 & 0.002173 & 0.006133 \\
    40 & 0.001146 & 0.000992 & 0.005687 & 0.001011 & 0.001062 & 0.002407 & 0.001039 & 0.002697 & \textbf{\textcolor{red}{0.974256}} & 0.009254 & 0.000449 \\
    41 & \textcolor{red}{0.967515} & 0.011243 & 0.002522 & 0.006031 & 0.007451 & 0.000945 & 0.00128 & 0.00105 & 0.001039 & 0.000506 & 0.000418 \\
    42 & 0.010232 & 0.00729 & 0.018894 & 0.11766 & 0.009483 & \textcolor{red}{0.892686} & 0.008646 & 0.007755 & 0.020424 & 0.008818 & 0.004006 \\
    \bottomrule
  \end{tabular}
  }
\end{table*}

\subsection{Masqueraders and Masquerade Detection}
Attacks can occur when an insider gains control of another employee's credentials with different privileges or when an outsider obtains a legitimate employee's credentials. Internal masqueraders often have greater organizational knowledge, making it easier to mimic employee behavior effectively. In contrast, external masqueraders may behave differently due to their lack of familiarity with the organization~\cite{khan2020database}. To address these threats, masquerade detection~\cite{bertacchini2008survey} serves as a specialized form of anomaly detection. In the context of SQL, external masquerade detection aims to identify out-of-scope queries that are not typical for database users. On the other hand, internal masquerade detection differentiates between a legitimate user's normal activities and a suspicious action indicative of a masquerader. Early methods for masquerade detection utilized traditional machine learning techniques such as Naive Bayes~\cite{maxion2002masquerade, maxion2003masquerade, wang2003one}, Support Vector Machines (SVMs)~\cite{wang2003one, kim2005empirical}, and K-Nearest Neighbors (KNN)~\cite{reciogarcia2023case, tarnowska2021log}. More recently, deep learning approaches~\cite{yuan2021deep, elmasry2018deep}, including RNN~\cite{yuan2018insider,yuan2019insider}, LSTM~\cite{azeezat2021conceptual}, and Bidirectional Long Short-Term Memory (Bi-LSTM)~\cite{hao2019bl, yu2018attention}, have been employed in masquerade detection, significantly enhancing detection accuracy. However, these masquerade detection methods are not well-suited to detect various types of SQL anomalies, such as data leaks, SQL injection (SQLi), and data sabotage. Additionally, they do not target the attack source, whether it originates externally or internally.  

\begin{figure}[t]
  \centering
  \includegraphics[width=0.6\linewidth]{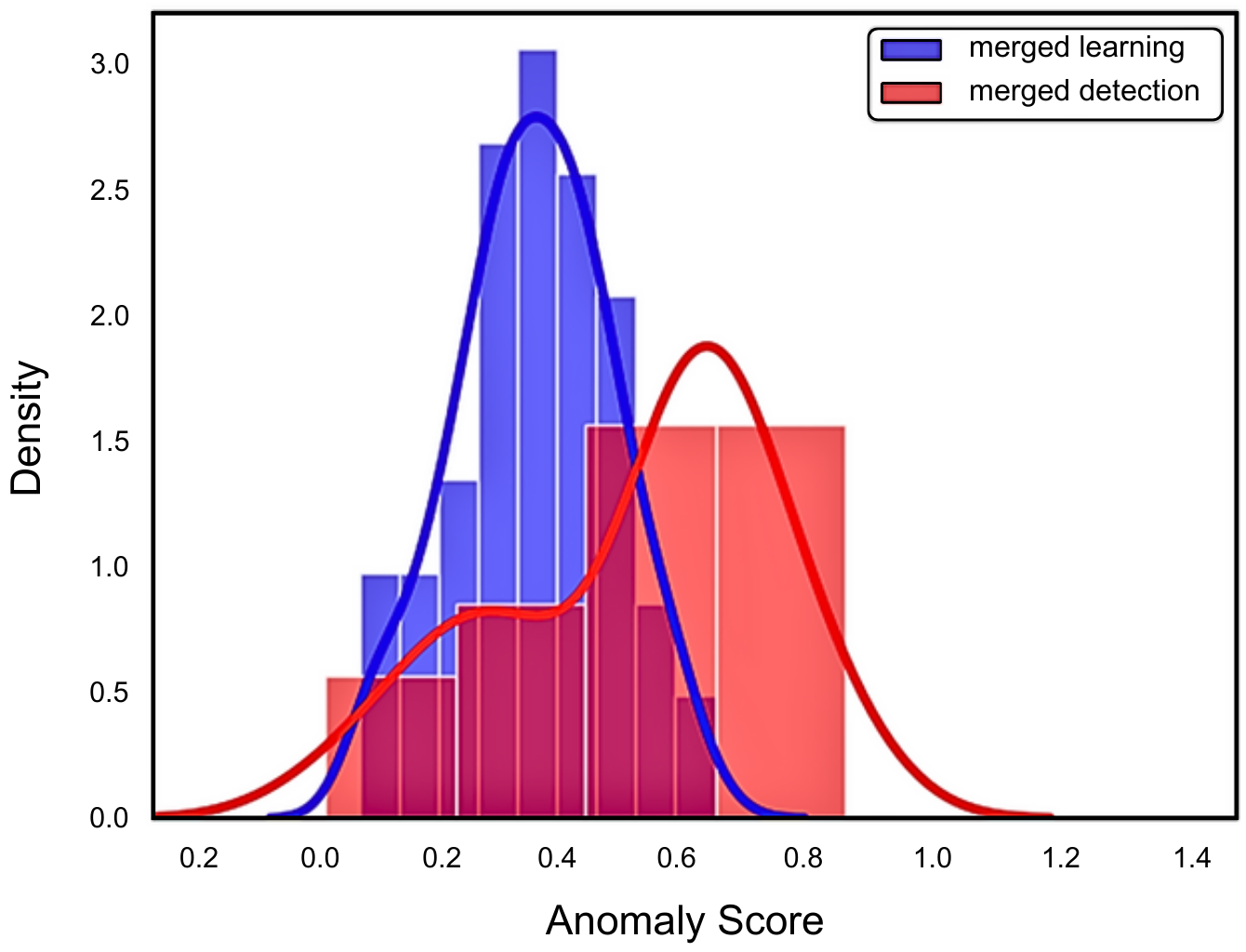}
  \caption{Distribution of averaged anomaly scores, normal (blue) and normal + abnormal out-of-scope queries (red).}
  \label{Fig4}
\end{figure}

\begin{table}[b] 
    \centering
    \caption{Out-of-scope queries examples: Data leaks, Data sabotage, and SQL injection} 
    \label{Table2}
    \resizebox{0.7\textwidth}{!}{ 
        \begin{tabular}{ll}
            \toprule
           Data leak & \texttt{select sensitive\_c1, sensitive\_c2 from T1} \\
            \midrule
            Data sabotage & \texttt{DROP TABLE T3} \\
                          & \texttt{UPDATE T1 SET COL1 = ? WHERE COL2 = ?} \\
            \midrule
            SQL injection & \texttt{SELECT * FROM T1 WHERE COL1 = ? OR ? = ?} \\
                          & \texttt{SELECT * FROM T1 WHERE COL1 = ? AND COL2 = ? OR ? = ?} \\
            \bottomrule
        \end{tabular}
    }
\end{table}

\section{Data}
\subsection{Data generation}
We generated SQL data from 2 datasets:
\begin{itemize}
    \item Short sequence dataset - we simulated a relational database with 3 user groups (HR, Finance, and DBA), each having approximately 100 unique normalized SQL queries. This dataset comprises 9 tables, 6 views, and ~50 attributes, with an average query length of 12 tokens. A unique challenge of our simulated data is the overlap between the different user regions, as multiple groups access the same tables and columns. In addition, the Finance group has access to sensitive data, whereas the other user groups have limited access to sensitive data columns.
    \item Long sequence dataset – We used an open source, comprehensive Customer Relationship Management (CRM) system built on PostgreSQL, installed and configured locally. This dataset includes 3 user groups, with approximately unique normalized 500 SQL per user. The average query length is 200 tokens, with the longest queries reaching up to 1900 tokens.
\end{itemize}

\subsection{Data preprocessing}
The following stages were followed in data preprocessing:
\subsubsection{Data normalization and cleaning:} 
\begin{itemize}
\item SQL queries were converted to their normalized form by replacing literal values with question marks '?', allowing the model to learn the basic form of the query without focusing on variable values.
\item All queries were converted to lowercase.
\item A fixed number of spaces was maintained between the tokens to create a uniform pattern.
\item Duplicate queries are removed for each user to prevent the model from being biased toward frequent queries.
\end{itemize}

\subsubsection{Overcome model embedding vector capacity limit}
For long queries that produce extensive input vectors (after tokenization), we split the input into several vectors based on the model embedding capacity (e.g., 512 tokens for BERT and 1024 tokens for LLaMA). We then averaged the embedding vector chunks for each query based on the assumption that using the element-wise sum or mean of the word embeddings across all words in the sentence effectively preserves the encoded meaning~\cite{white2015well}.

\section{Threat Model}
To contextualize our detection performance, we define two primary adversary profiles targeting database systems:

\textit{External Adversary (Outsider Threats):}
\begin{itemize}
    \item Access Level: Entry via compromised credentials, network exploits, or misconfigured public interfaces.
    \item Behavior Patterns: Lack historical behavioral alignment with authorized users. Mimicking legitimate SQL queries is difficult for attackers, especially external ones, as it requires access to application code or risky trial-and-error behavior that tends to trigger anomaly detectors. Tend to issue queries outside normal operational scope, or SQL injection.
    \item Attack Goals: Data leakage, or destructive actions such as table dropping or schema sabotage.
    \item Detection Strategy: Unsupervised learning using ensemble anomaly detectors that flag embedding vectors distant from learned normal patterns.    
\end{itemize}

\textit{Internal Adversary (Insider Threats):}
\begin{itemize}
    \item Access Level: Legitimate access with elevated privileges or stolen internal credentials.
    \item Behavior Patterns: Sophisticated mimicry of legitimate user behavior, different behavioral profile, often within expected schema and permission boundaries.
    \item Attack Goals: Subtle data exfiltration, privilege misuse, unauthorized report generation.
    \item Detection Strategy: Supervised learning using user-specific DistilBERT embeddings with probabilistic thresholds to catch behavior-role mismatches.
\end{itemize}

\section{Method}

\begin{table}[b] 
   \centering
   \caption{Parameter setting for each model - BERT, LLaMA, DistilBERT, LSTM, BiLSTM, SetFit}
  \label{Table3}
  \resizebox{0.7\textwidth}{!}{ 
    \begin{tabular}{llc}
        \toprule
         Model & Parameter & Selected Value \\
         \midrule
         BERT/LLaMA/DistilBERT & Number of iterations & 20 \\
                          & Batch size & 16 \\
                          & Number of epochs & 6 \\
                          & Learning rate & 1e-5 \\
                          & Activation & Softmax \\
         \midrule
         LSTM/Bi\_LSTM & Embedding\_dim & The average sequence length \\
                  & Batch size & 16 \\
                  & Number of epochs & 6 \\
                  & Activation & Softmax \\
         \midrule
        SetFit & Number of iterations & 20 \\
           & Batch size & 16 \\
           & Number of epochs & 1 \\
           & Learning rate & 1e-5 \\
    \bottomrule
  \end{tabular}
  }
\end{table}

Deep learning-based anomaly detection can be categorized into 3 types based on label availability: supervised, semi-supervised, and unsupervised deep anomaly detection~\cite{chalapathy2019survey}. Our approach utilizes supervised and unsupervised methods to detect 2 forms of masquerade attacks: external and internal. Supervised methods are used where both normal and abnormal data are present, enabling binary or multi-class classification~\cite{chalapathy2016bilstm, chalapathy2016investigation}. In contrast, unsupervised methods are employed when no labeled data is available or when anomalies need to be detected based on the internal properties of the data samples~\cite{hendrycks2018deep, su2019robust, tuor2017deep}. 
We used an unsupervised ensemble anomaly detector based on the fine-tuned DistilBERT model to detect out-of-scope queries. These queries, whose embeddings are significantly distant from the established embedding domain, are probably associated with external masqueraders, as their lack of familiarity with typical employee behaviors often results in anomalous activity~\cite{khan2020database}. 
We also evaluated several models for the supervised approach using labeled data. This approach may effectively detect internal masqueraders that can imitate the behavior of other employees within the organization~\cite{khan2020database}. We conducted distinct learning and detection periods for both methods.
The learning period refers to a designated time frame when users are assumed to perform only regular, non-malicious queries. During this period, the system collects user queries, which are pre-processed and used to train the models. Once the learning period concludes, the detection period begins. In this phase, the system monitors new queries performed on the database and alerts any abnormal queries. A detailed description is provided in the following sections.

\begin{figure}
  \centering
  \includegraphics[width=0.5\linewidth]{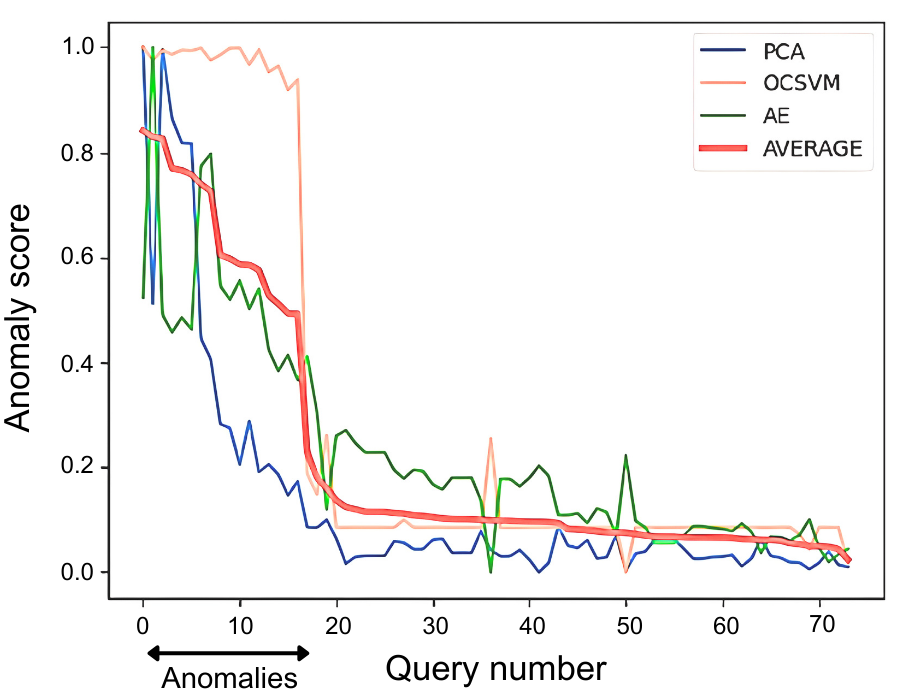}
  \caption{3 outlier detectors normal + anomalies scores sorted by average score, long sequences.}
  \label{Fig5}
\end{figure}

\begin{table}[b] 
    \caption{Evaluation results of 6 supervised models with different training sample sizes for long sequences dataset.}
    \label{Table4}
    \centering
    \setlength{\tabcolsep}{1pt} 
    \resizebox{0.7\textwidth}{!}{ 
        \begin{tabular}{lccccccccc}
            \toprule
            \multirow{2}{*}{\makecell{Sample \\ Size}} & \multicolumn{3}{c}{Fine-tuned DistilBERT} & \multicolumn{3}{c}{Fine-tuned BERT} & \multicolumn{3}{c}{Fine-tuned LLaMA} \\
            \cmidrule(lr){2-4} \cmidrule(lr){5-7} \cmidrule(lr){8-10}
            & Precision & Recall & F1 Score & Precision & Recall & F1 Score & Precision & Recall & F1 Score \\
            \midrule
            60K  & 0.34 & 0.43 & 0.38 & 0.42 & 0.57 & 0.48 & 0.14 & 0.38 & 0.20 \\
            240K & 0.66 & 0.50 & 0.57 & 0.75 & 0.62 & 0.68 & 0.42 & 0.29 & 0.34 \\
            390K & 0.87 & 0.65 & 0.74 & 0.82 & 0.71 & 0.76 & 0.63 & 0.47 & 0.54 \\
            540K & 0.88 & 0.86 & 0.87 & 0.96 & 0.79 & 0.87 & 0.81 & 0.75 & 0.78 \\
            660K & 0.96 & 0.94 & \textbf{0.95} & 0.90 & 0.93 & 0.91 & 0.94 & 0.91 & 0.92 \\
            \bottomrule
        \end{tabular}
    }
\end{table}

\subsection{Unsupervised part for out-of-scope query detection}

This part's primary objective is to detect distant vectors in the multidimensional vector embedding space. Such vectors are likely to fall outside the typical query domain of regular database users, indicating potential anomalous or out-of-scope queries.
DistilBERT~\cite{sanh2019distilbert} was chosen for its strong performance and reduced computational requirements. Using a Masked Language Model (MLM) randomly masks tokens in the input and predicts their meaning based on context, effectively capturing the inherent structure and dependencies within the SQL sequences~\cite{devlin2019bert}.
SQL is similar to human language, making it suitable for natural language processing techniques. In our unsupervised approach, we fine-tuned DistilBERT on SQL datasets, focusing first on queries collected during the learning period. We then extract the last hidden state embedding, resulting in a 768-dimensional representation. These embeddings serve as an input for creating an ensemble anomaly detector by applying 3 outlier detectors. The 3 detectors trained on the SQL embeddings are Principle Component Analysis (PCA)~\cite{shyu2003novel}, Autoencoders (AE)\cite{baldi1989neural}, and One-Class Support Vector Machine (OCSVM)~\cite{alaverdyan2020regularized}. Reconstruction errors were calculated for both the AE and PCA models. The embeddings extracted from PCA dimensionality reduction~\cite{hasan2021review}, which preserves 98\% of the common variance, were used as input for the OCSVM, with decision scores subsequently normalized. The final anomaly score was determined by averaging all 3 normalized scores. This score measured the deviation of an SQL query from the overall set. Out-of-scope queries, including data leaks, attacks such as SQLi attacks, and data sabotage, typically receive the highest anomaly scores. A threshold (with a confidence interval) determines the upper limit of our normal learning period queries.
The entire process is repeated on the SQL queries from the detection period, using the threshold set during the learning period to detect out-of-scope queries.

\begin{figure}
  \centering
     \setlength{\tabcolsep}{1pt} 
  \includegraphics[width=0.5\linewidth]{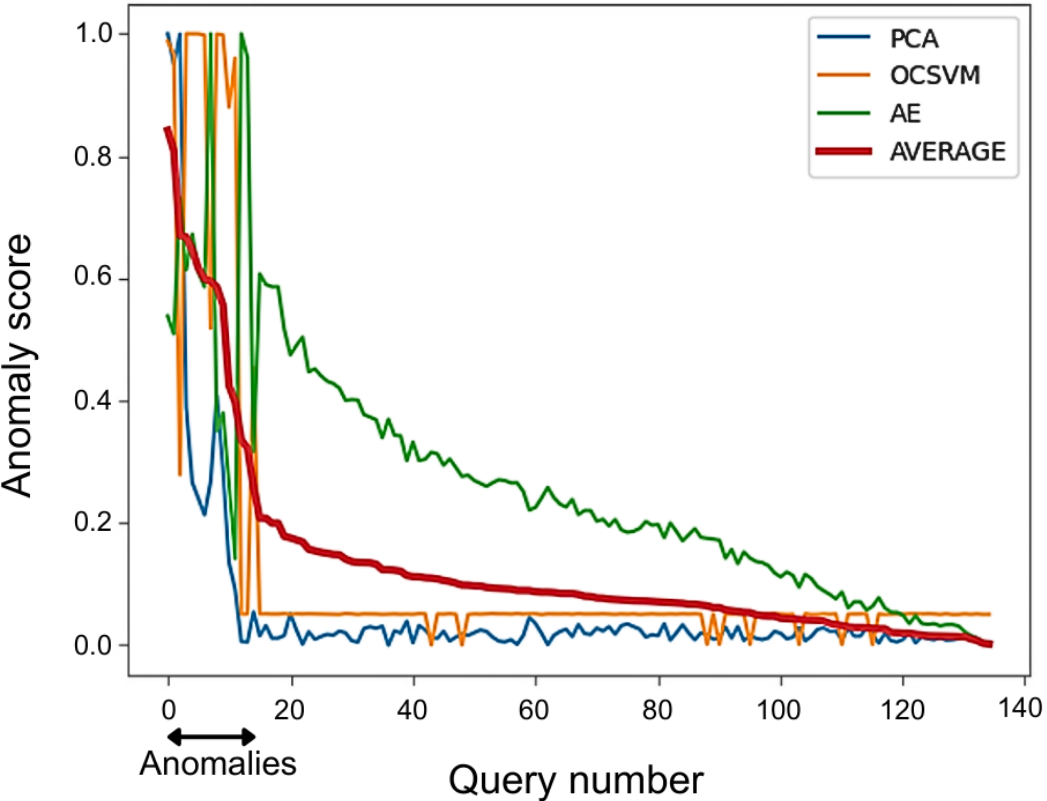}
  \caption{3 outlier detectors normal + anomalies scores sorted by average score, short sequences.}
  \label{Fig6}
\end{figure}

\begin{table}[t] 
    \caption{Evaluation results of 3 supervised LSTM, Bi-LSTM, and SetFit models with different training sample sizes for long sequences dataset.}
    \label{Table5}
    \centering
    \setlength{\tabcolsep}{1pt} 
    \resizebox{0.7\textwidth}{!}{ 
        \begin{tabular}{lccccccccc}
            \toprule
            \multirow{2}{*}{\makecell{Sample \\ Size}} & \multicolumn{3}{c}{LSTM} & \multicolumn{3}{c}{Bi-LSTM} & \multicolumn{3}{c}{SetFit} \\
            \cmidrule(lr){2-4} \cmidrule(lr){5-7} \cmidrule(lr){8-10}
            & Precision & Recall & F1 Score & Precision & Recall & F1 Score & Precision & Recall & F1 Score \\
            \midrule
            60K  & 0.22 & 0.11 & 0.15 & 0.22 & 0.35 & 0.27 & 0.58 & 0.65 & 0.61 \\
            240K & 0.30 & 0.11 & 0.16 & 0.59 & 0.38 & 0.46 & 0.80 & 0.65 & 0.72 \\
            390K & 0.22 & 0.12 & 0.16 & 0.77 & 0.60 & 0.67 & 0.90 & 0.74 & 0.81 \\
            540K & 0.36 & 0.125 & 0.18 & 0.90 & 0.78 & 0.84 & 0.92 & 0.84 & 0.88 \\
            660K & 0.35 & 0.13 & 0.19 & 0.90 & 0.93 & 0.91 & 0.97 & 0.98 & \textbf{0.97} \\
            \bottomrule
        \end{tabular}
    }
\end{table}

\subsection{Supervised part for in-scope query detection}

\begin{figure}[b]
  \centering
  \includegraphics[width=0.5\columnwidth]{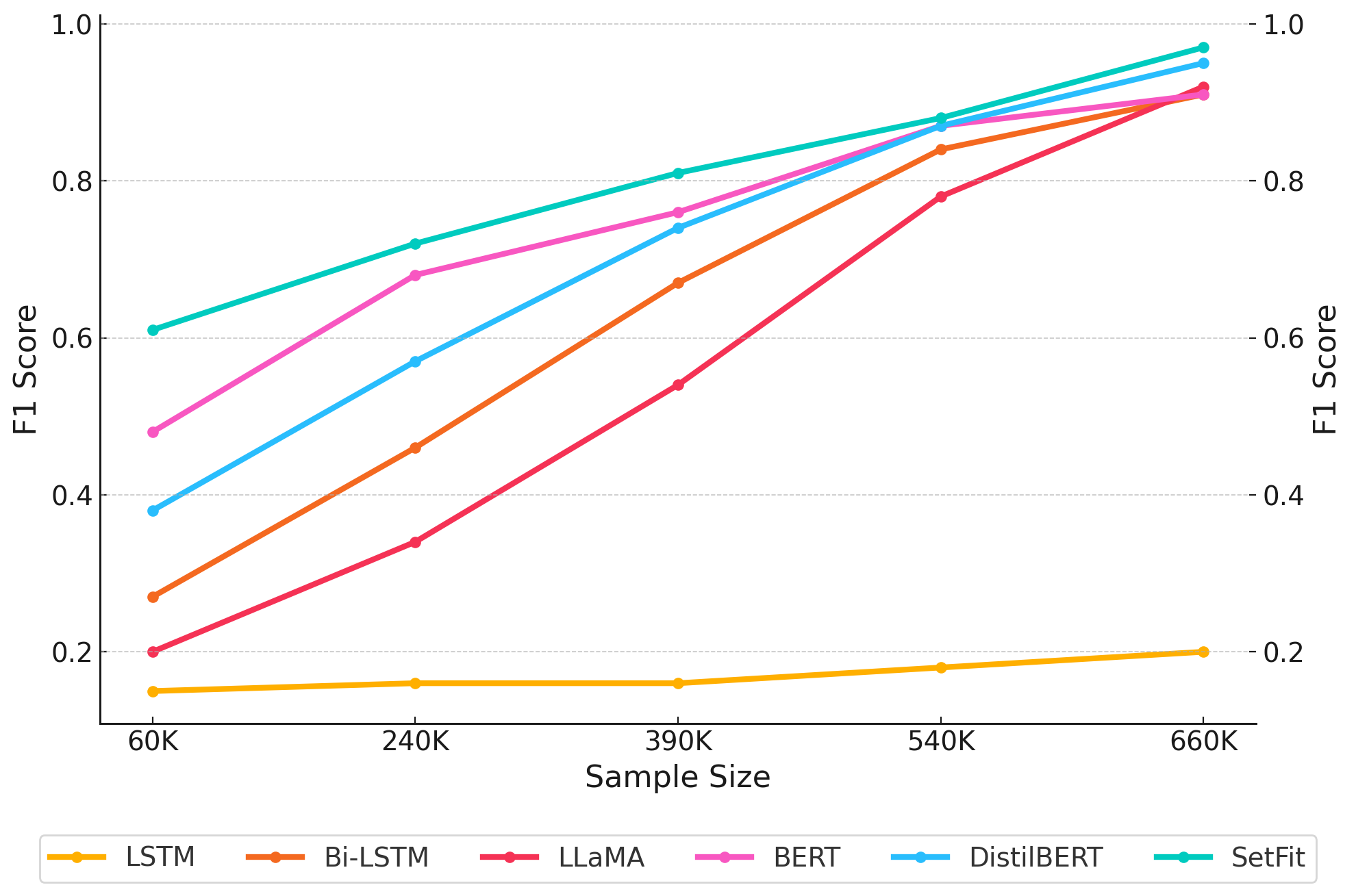}
  \caption{F1 scores of 6 supervised models with different training sizes for long sequences dataset.}
  \label{Fig7}
\end{figure}

In the supervised part of our study, the learning period serves as a phase in which a probabilistic classifier models the behavior of each role or user and assigns labels for them (for example, in our short sequence dataset – Finance, HR, and DBA are the user groups). A threshold probability is then determined for each role or user. These probabilities represent the likelihood that a given query belongs to a specific user based on the classifier's learned patterns. This learning period contains several key stages, as outlined in the scheme illustrated in Figure~\ref{Fig2}.
The validation data set generates a probability matrix (Table~\ref{Table1}). Each column corresponds to one of the 11 users (labeled from 0 to 10) and is associated with certain validation queries. Each row represents a set of user probabilities calculated by the probabilistic classifier for a given validation input instance (validation query) within a stratified validation set. The probabilities across each row are normalized, ensuring their sum equals 1. Each user's probability threshold is determined based on the classification results and the validation input dataset. The lowest significant probability value is determined as the probability threshold. During the detection period, queries are tagged as ‘Normal’ or ‘Abnormal’ according to the probability matrix produced by the classification layer.

The highest probability for the first validation query is 0.956105, associated with user 2 (first row, column 2 in Table~\ref{Table1}). Similarly, the highest probability for a second validation query is 0.936215 and is associated with user 5 (second row, column 5 in Table~\ref{Table1}). This approach identifies the highest probability value for each validation query as the significant probability. In Table~\ref{Table1}, these significant values are marked in bold red.
The second step determines the lowest significant probability value for each user. For example, for user 2, the lowest significant probability value is 0.867791 (row 23, column 2 in Table~\ref{Table1}). Similarly, for user 5, the lowest significant probability value is 0.892686 (row 42, column 5 in Table~\ref{Table1}). These lowest significant probability values are then used to establish a respective probability threshold for each user. In the detection period, a detection input instance based on a query enters a data repository like a database and is classified using the trained probabilistic classifier. The classifier assigns a probability indicating the likelihood that the query belongs to a certain user. This probability is then used to determine if the detection query is abnormal. The detection period contains several key stages described in the scheme illustrated in Figure~\ref{Fig3}.
During the detection period, anomalies in the labeled datasets can be detected in 2 ways. First, an anomaly is flagged if the highest probability is associated with another user. Second, an anomaly is identified if the highest probability belongs to the current user but falls below the threshold (with confidence interval) established during the learning phase. We used all labeled data (SQL queries labeled by each user) to fine-tune various models, including the RNN models (LSTM, Bi-LSTM) and the LLM models (BERT, DistilBERT, LLaMA). Additionally, we fine-tuned the Sentence Transformer Fine-tuning (SetFit)~\cite{tunstall2022efficient}, a novel approach to few-shot text classification. SetFit is significantly faster in inference and training compared to similar methods like T-FEW~\cite{liu2022few}, ADAPET~\cite{tam2021improving}, and PERFECT~\cite{mahabadi2022perfect} while also delivering strong performance with significantly smaller and more efficient base models.

\begin{figure}[t]
  \centering
  \includegraphics[width=0.5\columnwidth]{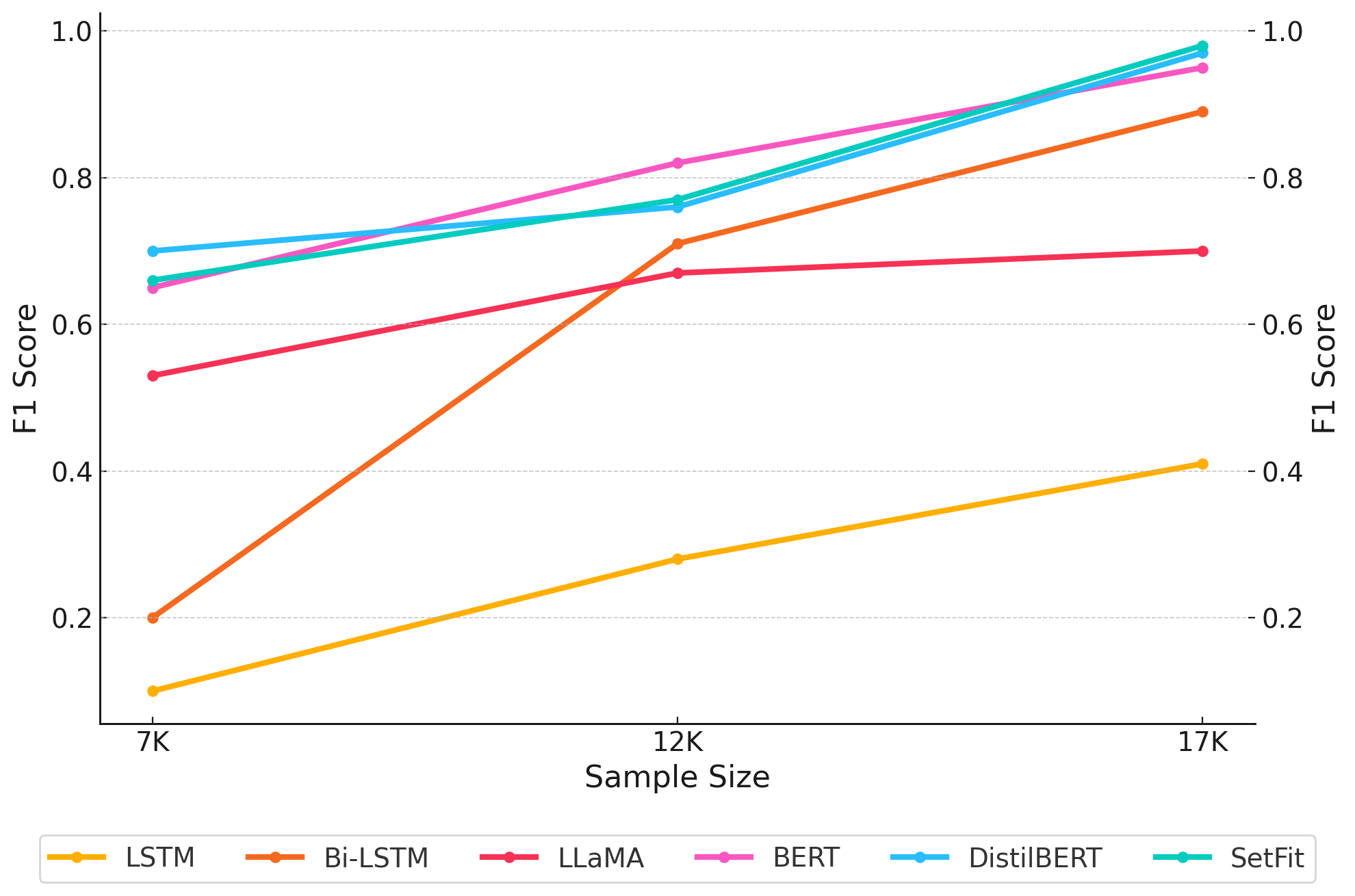}
  \caption{F1 scores of 6 supervised models with different training sizes for short sequences dataset.}
  \label{Fig8}
\end{figure}

\section{Results}
The following sections present the results of unsupervised and supervised anomaly detection methods applied to SQL. We first evaluate the unsupervised model, which uses fine-tuned DistilBERT embeddings alongside the ensemble anomaly detectors on unlabeled data. We then assess the performance of the supervised model using labeled SQL queries.

\subsection{Unsupervised, out-of-scope query detection results}

We conducted several analyses to evaluate the unsupervised model and gain insights into its performance. These include:
\begin{itemize}
    \item Visualizing the distributions of anomaly scores compared to normal scores of embedding vectors to recognize the difference between them.
    \item Comparing the anomaly scores generated by the 3 outlier detectors and their averaged scores across the long and short sequences datasets.
\end{itemize}

Table~\ref{Table2} presents out-of-scope SQL query examples, including those that could result in data leaks, data sabotage, or SQLi. 
In anomaly or outlier detection~\cite{ruff2021unifying, chalapathy2019survey}, it is generally assumed that only a small number of anomalous data instances exist in the distribution. Figure~\ref{Fig4} shows the distribution of anomaly scores from the learning period (composed only of normal queries) compared to the detection period (consisting of normal and abnormal queries). These scores are derived from the combined output of the 3 normalized outlier detectors. The distribution appears to follow a near-normal pattern, with most SQL queries clustered around the mean, indicating typical anomaly scores. However, the distribution also reveals the presence of some outlier SQLs that exhibit significantly higher anomaly scores, marking them as potential anomalies.
The anomaly scores from the 3 outlier detectors for the long and short SQL sequences consisting of normal and abnormal queries during the detection period are presented in Figures~\ref{Fig5} and ~\ref{Fig6} (respectively). The red line represents the average of those 3 detectors, with scores sorted based on this average. Overall, the OCSVM tends to produce the highest anomaly scores, while the AE and PCA produce the lowest scores. However, in some cases, AE and PCA scores are more effective at revealing specific anomalies than OCSVM. Notably, in the short sequences graph (Figure~\ref{Fig6}), the AE scores appear more volatile than in the long sequences graph (Figure~\ref{Fig5}). Despite these differences, both graphs clearly distinguish between normal and abnormal queries when using the average of the 3 outlier detectors.

\begin{table}[b] 
    \caption{Evaluation results of 3 supervised Fine-tuned DistilBERT, BERT, and LLaMA models with different training sample sizes for short sequences dataset.}
    \label{Table6}
    \centering
    \setlength{\tabcolsep}{1pt} 
    \resizebox{0.7\textwidth}{!}{ 
        \begin{tabular}{lccccccccc}
            \toprule
            \multirow{2}{*}{\makecell{Sample \\ Size}} & \multicolumn{3}{c}{Fine-tuned DistilBERT} & \multicolumn{3}{c}{Fine-tuned BERT} & \multicolumn{3}{c}{Fine-tuned LLaMA} \\
            \cmidrule(lr){2-4} \cmidrule(lr){5-7} \cmidrule(lr){8-10}
            & Precision & Recall & F1 Score & Precision & Recall & F1 Score & Precision & Recall & F1 Score \\
            \midrule
            7K  & 0.81 & 0.62 & 0.70 & 0.72 & 0.60 & 0.65 & 0.79 & 0.40 & 0.53 \\
            12K & 0.83 & 0.70 & 0.76 & 0.87 & 0.78 & 0.82 & 0.74 & 0.61 & 0.67 \\
            17K & 0.97 & 0.98 & \textbf{0.97} & 0.94 & 0.96 & 0.95 & 0.69 & 0.72 & 0.70 \\
            \bottomrule
        \end{tabular}
    }
\end{table}

\subsection{Supervised, in-scope query detection results}

The supervised component of our approach uses labeled SQL data corresponding to different database users. A query is deemed abnormal if it is assigned to a different user or if it is assigned to the correct user but falls below the established probability threshold. Conversely, a query is considered normal if it is classified to the same labeled user and exceeds the established threshold. The labeled dataset is divided into training and testing sets in an 85:15 ratio. To ensure balance, the training data includes an equal number of SQL queries for each user, which are then combined to create a few-shot training set.
Given the potential instability of evaluation results from a small-size training set, we conducted 5 experiments for each model and sample size per class. The parameter settings for each model are described in Table~\ref{Table3}. The F1 scores (2 * (precision * recall) / (precision + recall)) are presented in Figures~\ref{Fig7} and~\ref{Fig8} for LSTM, Bi-LSTM, LLaMA, BERT, DistilBERT, and SetFit models for long and short SQL sequences, respectively. The training data sizes were examined in the long sequence dataset (Figure~\ref{Fig7}): 60K, 240K, 390K, 540K, and 660K. The training data sizes examined in the short sequence dataset (Figure~\ref{Fig8}) are 7K, 12K, and 17K. 
The averaged precision, recall, and F1 scores for short and long-sequence datasets are described in Tables~\ref{Table4}, ~\ref{Table5}, ~\ref{Table6} and ~\ref{Table7}. While Tables~\ref{Table4} and ~\ref{Table5} describe the scores on long sequence datasets with training data sizes varying from 60K to 660K, Tables~\ref{Table6} and ~\ref{Table7} describe the scores on short sequence datasets with training data sizes varying from 7K to 17K.
As expected, the model’s performance improves as the sample size increases. In addition, across all sample sizes, we noticed that LLaMA performances are lower than BERT/DistilBERT when using our limited data. Our findings align with Bumgardner's conclusion~\cite{bumgardner2024local} that claims the BERT model achieves high performance on smaller datasets, whereas the LLaMA models excelled with the larger datasets. This discrepancy can be attributed to the simpler classification challenge of smaller datasets featuring fewer class labels and examples compared to the complexity of the larger dataset, which offers more complex training data. 
Another significant observation is that the Bi-LSTM models consistently outperformed LSTM models. This is because the Bi-LSTM processes input in both directions, leveraging contextual information from both sides, unlike LSTM, which processes data in a single direction. Additionally, the fine-tuned SetFit model yielded the best results when trained on the largest data sets, outperforming the fine-tuned DistilBERT with the same data size. Conversely, LSTM models produced the lowest performance, highlighting the advantages of using SetFit for fine-tuning pre-trained models when labeled data is limited. Overall, the experimental results demonstrate the effectiveness of creating a small set of manually labeled SQL, fine-tuning a pre-trained model with SetFit, and subsequently using it for automated SQL classification. We combined equal numbers of normal and abnormal sessions into a few-shot training set that resulted in improved performance as the number of samples per user increased.

\begin{table}[t] 
    \caption{Evaluation results of 3 supervised LSTM, Bi-LSTM, and SetFit models with different training sample sizes for short sequences dataset.}
    \label{Table7}
    \centering
    \setlength{\tabcolsep}{1pt} 
    \resizebox{0.7\textwidth}{!}{ 
        \begin{tabular}{lccccccccc}
            \toprule
            \multirow{2}{*}{\makecell{Sample \\ Size}} & \multicolumn{3}{c}{LSTM} & \multicolumn{3}{c}{Bi-LSTM} & \multicolumn{3}{c}{SetFit} \\
            \cmidrule(lr){2-4} \cmidrule(lr){5-7} \cmidrule(lr){8-10}
            & Precision & Recall & F1 Score & Precision & Recall & F1 Score & Precision & Recall & F1 Score \\
            \midrule
            7K  & 0.06  & 0.25  & 0.10  & 0.14  & 0.37  & 0.20  & 0.62  & 0.70  & 0.66 \\
            12K & 0.24  & 0.35  & 0.28  & 0.75  & 0.67  & 0.71  & 0.75  & 0.80  & 0.77 \\
            17K & 0.40  & 0.42  & 0.41  & 0.88  & 0.90  & 0.89  & 0.98  & 0.975 & \textbf{0.98} \\
            \bottomrule
        \end{tabular}
    }
\end{table}

\section{Conclusions}

Detecting abnormal database access behavior remains a critical challenge, especially given the limitations of existing approaches in processing complex statements, detecting abnormal behavior, performing root cause analysis, and maintaining precision. To address these challenges, we introduce an innovative approach that effectively detects abnormal database access behavior from both external and internal aspects. Our approach leverages advanced techniques for extracting semantic information from SQL operation statements and integrates 2 distinct detection techniques, unsupervised and supervised, to achieve highly accurate anomaly detection. Extensive experiments conducted on datasets from 2 different database scenarios indicate that our techniques outperform SOTA methods. The results show that our model enhances detection accuracy and provides a more comprehensive analysis of user behavior, making it well-suited for real-world applications. By refining anomaly detection with a blend of semantic understanding and advanced machine learning techniques, our approach sets a new standard for ensuring database security and integrity.

\bibliographystyle{unsrt}
\bibliography{references}  






\end{document}